\documentclass[sn-nature]{sn-jnl}

\usepackage{threeparttable}  
\usepackage{multicol}  
\usepackage{multirow} 
\usepackage{graphicx}%
\usepackage{multirow}%
\usepackage{amsmath,amssymb,amsfonts}%
\usepackage{amsthm}%
\usepackage{mathrsfs}%
\usepackage[title]{appendix}%
\usepackage{xcolor}%
\usepackage{textcomp}%
\usepackage{manyfoot}%
\usepackage{booktabs}%
\usepackage{algorithm}%
\usepackage{algorithmicx}%
\usepackage{algpseudocode}%
\usepackage{listings}%

\theoremstyle{thmstyleone}%
\theoremstyle{thmstyletwo}%

\theoremstyle{thmstylethree}%

\begin{document}

\title[Article Title]{Energy Internet: A Standardization-Based Blueprint Design}

\author*[1]{\fnm{Ye} \sur{Guo}}\email{guo-ye@sz.tsinghua.edu.cn}
\author[1]{\fnm{Hanyang} \sur{Lin}}\email{linhy22@mails.tsinghua.edu.cn}
\author[2]{\fnm{Hongbin} \sur{Sun}}\email{shb@tsinghua.edu.cn}

\affil[1]{\orgdiv{Tsinghua Berkeley Shenzhen Institute}, \orgname{Tsinghua University}, \orgaddress{\street{Lishui Road}, \city{Shenzhen}, \postcode{518000}, \state{Guangdong}, \country{China}}}

\affil[2]{\orgdiv{Department of Electrical Engineering}, \orgname{Tsinghua University}, \orgaddress{\street{West Main Building}, \city{Tsinghua Park}, \postcode{100084}, \state{Beijing}, \country{China}}}


\abstract{The decarbonization of power and energy systems faces a bottleneck: The enormous number of user-side resources cannot be properly managed and operated by centralized system operators, who used to send dispatch instructions only to a few large power plants. To break through, we need not only new devices and algorithms, but structural reforms of our energy systems. Taking the Internet as a paradigm, a practicable design of the Energy Internet is presented based on the principle of standardization. A combination of stylized data and energy delivery, referred to as a Block of Energy Exchange (BEE), is designed as the media to be communicated, which is parsed by the Energy Internet Card. Each Energy Internet Card is assigned a unique MAC address, defining a participant of the Energy Internet, whose standardized profile will be automatically updated according to BEE transfers without the intervention of any centralized operator. The structure of Energy Internet and protocols thereof to support the transfer of BEE are presented. System operators will become Energy Internet Service Providers, who operate the energy system by flow control and dispatching centralized resources, which is decoupled from users' behaviors in the Energy Internet. Example shows that the Energy Internet can not only reduce carbon emissions via interactions between peers, but also promotes energy democracy and dwindles the gap in energy equity.}


%

\keywords{Energy Internet, Internet, Power and Energy Systems, Internet Protocols, Multi-energy Systems}



\maketitle

\newpage

\section{Introduction}\label{sec1}

For decades, our electric power systems have been operated in a centralized manner \cite{act2005energy}. A few large thermal power plants, scheduled by the system operator either in a vertically integrated manner or through a wholesale electricity market, dominate the system operation \cite{yang2017small}. Energy, in general, flows unidirectionally: generated from thermal units, traverses the high-voltage transmission and low-voltage distribution networks, then finally arrives at consumers \cite{weo}. Many other energy sectors, such as natural gas and heat, share similar structures \cite{6geidl2006energy}.

\begin{figure}[H]
	\centering
	\includegraphics[width=5.0in]{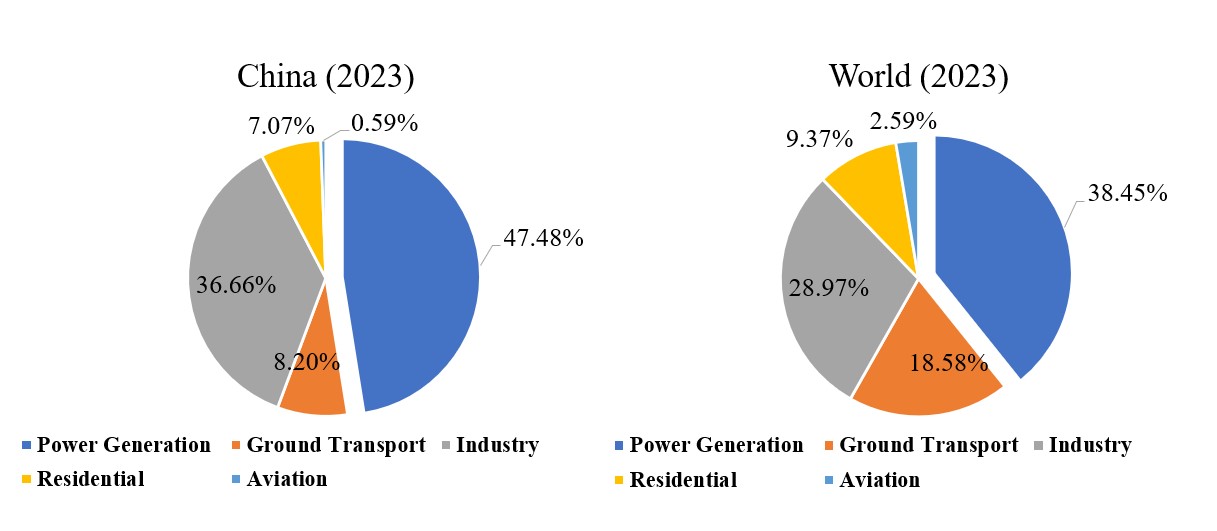}
	\caption{Carbon Emission by Sectors in China and the World\cite{WinNT}.}
	\label{fig:Carbon}
\end{figure}

Fundamental changes have happened, however, driven by the global efforts towards carbon neutrality \cite{kammen2016city}. Energy generation accounts for the largest portion of carbon emissions among all sectors globally \cite{liu2024global}, as shown in Fig. \ref{fig:Carbon}. Thus, transitioning from fossil-fuelled energy generation to renewable energy sources is a paramount and indispensable task for sustainability \cite{wang2023accelerating}. On the one hand, wind and solar units are much more suitable to be deployed in a distributed manner than thermal units \cite{helveston2022quantifying,9sgouridis2019comparative,carlino2023declining}. On the other hand, to accommodate uncertainties of renewable energies, our energy system has been integrating an enormous number of user-side resources, such as batteries \cite{cheema2024giant,wang2014lithium}, electric vehicles \cite{27liu2024transforming,10liu2019challenges}, flexible load resources \cite{21creutzig2024demand}, which are again distributed assets \cite{14thomas2019general} .

Challenges arise, consequently, since conventional centralized operation mode may not be compatible with new characteristics of our energy systems in the carbon-neutrality era\cite{2debnath2018challenges}: Multi-entities, distributed resources \cite{31kittner2017energy,19smith2022effect}, and diversified energy products considering uncertainties \cite{23anadon2017integrating}, carbon profiles \cite{20fan2023net}, and inter-temporal operational limits of resources \cite{guo2021pricing}. An immediate problem on the table is: distributed energy resources (DERs) are too many, too small, and too random \cite{17heptonstall2021systematic}, making them scarcely possible to be properly managed by the system operator \cite{veers2019grand}, who used to send instructions only to a few large power plants \cite{29sturmer2024increasing}.

Multiple technical solutions have been studied, including virtual power plants (VPP) and Peer-to-Peer (P2P) trading platforms. The former consolidated diverse types of DERs into a cohesive entity to interact with the power grid \cite{pudjianto2007virtual,7063236}, while the latter allows users to trade excessive energy with their neighbors under the approval of the system operator \cite{18irena2020innovation, 15pena2022integration}. They are both important and novel techniques for energy system operations with enormous distributed resources \cite{11parag2016electricity, 13morstyn2018using}. However, they focus on particular aspects but do not provide us with a holistic picture of future energy systems. In light of the fundamental and structural changes in resources, our energy system needs not only new devices, platforms, and algorithms, but also a brand-new architecture tailored for multi-entity distributed energy resources. 

To that end, scholars have noticed an excellent paradigm -- The Internet \cite{theeco}. In the 1990s, also known as the Web 1.0 era \cite{peterson2007computer}, the Internet was in a similar shape as energy systems: A few web portals dominate information flow, users are by and large read-only, and most information flows unidirectionally \cite{watts1998collective}. Nowadays, however, the Internet has become an open system connecting billions of users who actively communicate with each other \cite{hilbert2011world}. These transitions are exactly what we want in energy systems, thus the notion of Energy Internet has been extensively discussed \cite{3huang2010future,8sun2018integrated,16wang2017survey}.

Unfortunately, given all the efforts above, energy systems today are still heavily centralized \cite{2021neo,24zhang2022assessing}. This is in sharp contrast to the Internet, which evolved from Web 1.0 to the current Web 2.0 era in about a decade. In fact, we cannot even put forth a clear definition of Energy Internet, but are stagnating at a relatively vague level: It is an Internet-like energy system \cite{3huang2010future}, it should be more open, but energy is still different from information, etc \cite{25wang2022cyber,26liu2021design}. 

This is exactly what we want to address in this article. We first depict the basic setting of the Energy Internet: What is the media to be exchanged and who are participants. Subsequently, a holistic architecture of the Energy Internet is presented, closely following that of the Internet. Taking the TCP/IP model as a basis, energy exchange protocols are designed in this paper, which we believe will break through bottlenecks towards distributed energy exchange without any involvement from the system operator. Energy Internet applications and how energy systems should be operated are discussed. Finally, a toy example is given to illustrate how an Energy Internet operates and how it differs from the traditional structure. 

\section{Block of Energy Exchange (BEE)}

Internet is constructed based on the standardization of data -- information and data are organized in a binary sequence--a byte, which is communicated between users.

\begin{figure}[H]
	\centering
	\includegraphics[width=5.0in]{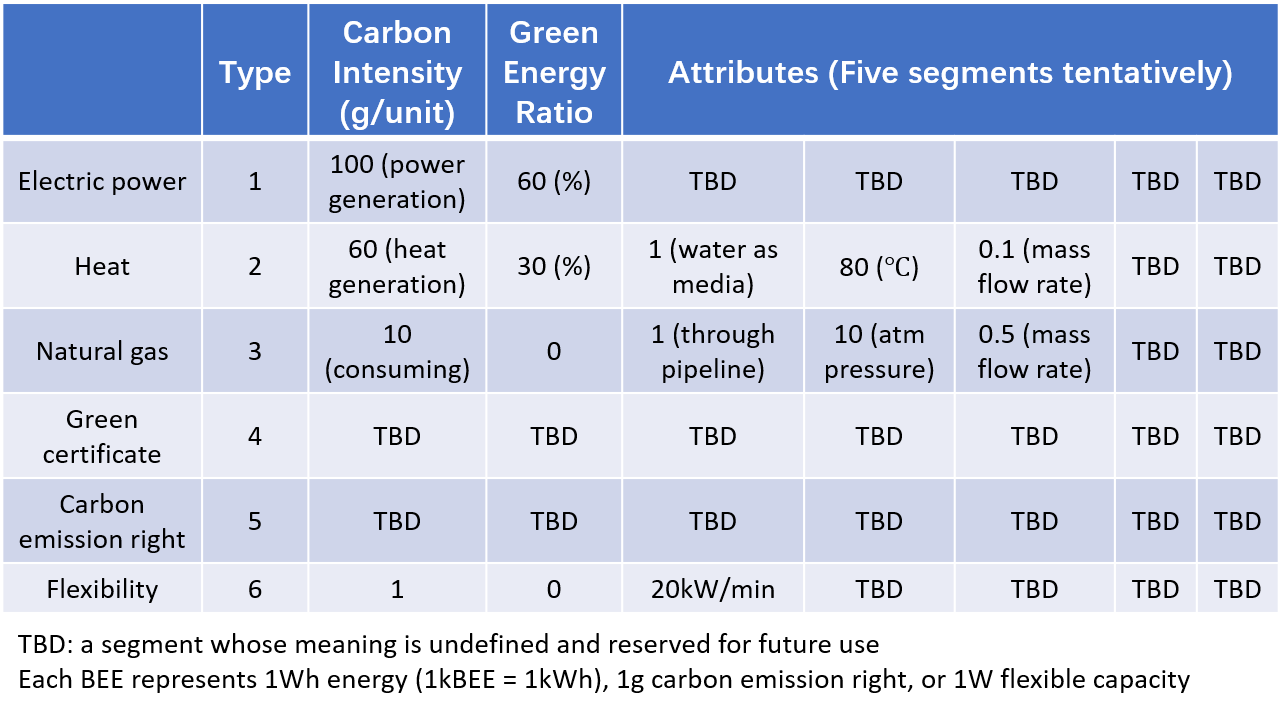}
	\caption{Format of a Block of Energy Exchange (BEE).}
	\label{fig:Energy}
\end{figure}

Traditionally, electric power is considered homogenous goods \cite{28yan2019city}. Things may change, however, considering their carbon intensity and portions of green energy. For other types of energy, such as heat, key factors may include media, grade, and mass flow rate \cite{7mohammadi2017energy}. Here, we define a Block of Energy Exchange (BEE), which is a combination of an array of data in the ex-ante stage and associated delivery of energy-related products in the real-time stage. The format of its data part is designed as a nine-entry array and is illustrated in Fig. \ref{fig:Energy}.

BEEs will be the objects exchanged in the energy internet. 

\section{Resources in the Energy Internet}


For Energy Internet resources, we propose to establish a profile record for each of them. In this paper, we tentatively design the profile of an Energy Internet user as:

\begin{figure}[H]
	\centering
	\includegraphics[width=4.5in]{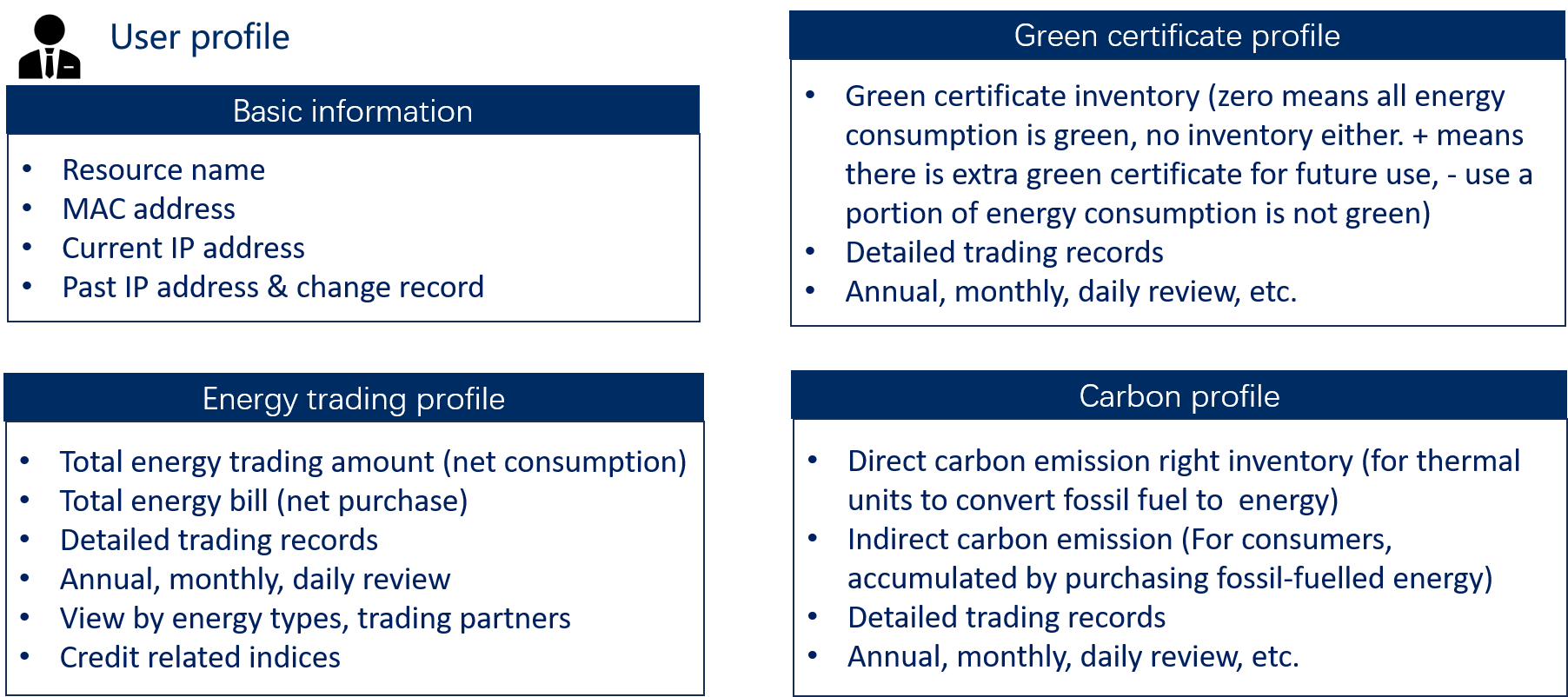}
	\caption{User profile for resources in the Energy Internet.}
	\label{fig:user}
\end{figure}

Essentially, when BEEs are transferred in Energy Internet, client profiles will be automatically updated, including their energy trading records and inventories of green certificate and carbon emission rights, without any involvement from the system operator. Techniques of Blockchain, Smart Contract, and Automatic Billing can be employed to better boost a distributed and privacy-preserved ledger for Energy Internet clients, but details are beyond the scope of this paper.

We also propose the notion of ``Energy Internet Card'', assembling to the Internet Card, being an integrated hardware and software module that can parse the BEE transferred in the Energy Internet and automatically update profiles of Energy Internet clients. Each Energy Internet Card, associated with a MAC address, defines a resource or user of the Energy Internet.

\section{Energy Internet Structure: A bird eye's view}

The structure of Energy Internet is designed in Fig. \ref{fig:EI}, which closely follows that of the Internet (ref) with counterparts of all components defined as follows. 

\begin{figure}[H]
	\centering
	\includegraphics[width=5in]{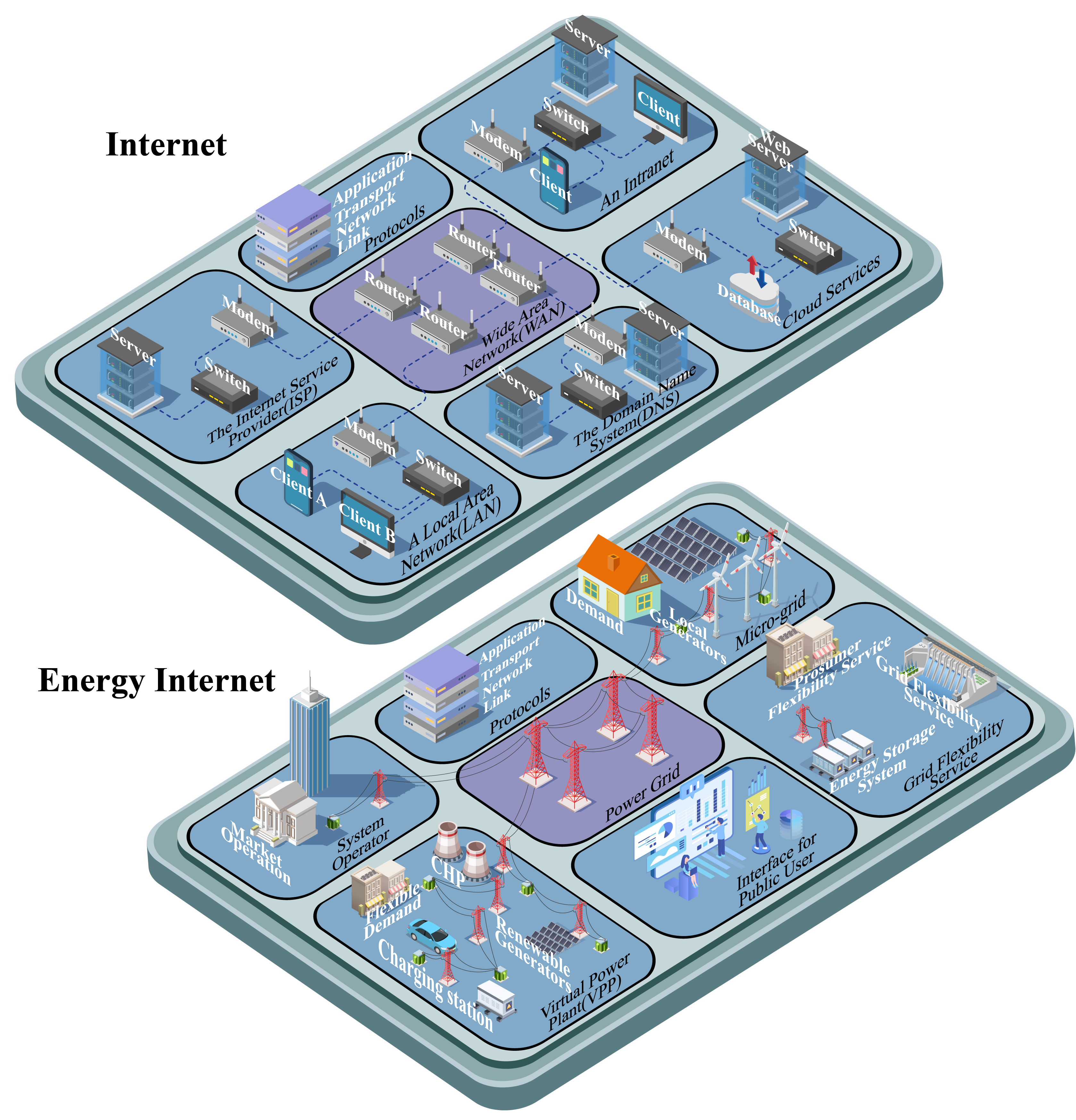}
	\caption{Energy Internet: A bird eye's view.}
	\label{fig:EI}
\end{figure}

\begin{itemize}
  \item Energy Local Area Network (LAN) -- Virtual power plants: Similar to the Internet, Energy LANs are basic units of the Energy Internet. An Energy LAN is a collection of energy resources in proximity, enabling local P2P energy trading and interacting with the main grid or other Energy LANs. In this paper, we use the term ``virtual power plants'' to include all possible forms of Energy LANs above, which may also with the name of multi-energy industrial parks, rural energy ecosystems, or electric vehicle aggregators. 
  \item Energy Intranet -- Microgrids: A special form of Energy LAN that is able to disconnect with the rest of the Energy Internet, which is perfectly in line with the notion of microgrids.
  \item Energy Wide Area Network (WAN) -- Interconnected power transmission grid: Similar to the Internet, connecting numerous Energy LANs located in a wide range of area together further leads to the Energy WAN. VPP operators will serve as the Energy Internet router, defining Energy LAN IP address and is associated with an Energy WAN IP address. The most important form of Energy WAN is the interconnected power transmission grid, while in some cases natural gas systems may also become a part.
  \item Energy Internet Service Provider (ISP)--System operators: In the Energy Internet, current system operators will evolve into Energy Internet Service Providers, who dispatch centralized resources and implement quantity control to provide energy prosumers with reliable Energy Internet Services. Most existing functions in system operators can be persevered. 
  \item Energy Web Servers -- Ancillary services provided by prosumers: In supplement to centralized services from Energy Internet Service Providers, there are also Energy Internet Services from prosumers, including reserve, regulation, carbon footprint offset, green certificate purchase, etc. 
  \item Database Servers -- Storage services: In the Energy Internet, each BEE may be heterogenous and may be stored in energy storage systems with its owner labelled -- similar to cloud database in the Internet. 
  \item Energy Internet Domain Name System (DNS) -- Interface for public users: In the Internet, the domain name system transfers IP addresses to user-friendly website domains. A similar role may appear in future Energy Internet, where each resource is also deployed an Energy MAC address, an Energy IP address, and a resource name for public clients.
\end{itemize}

\section{Energy Internet Protocols}

\begin{figure}[H]
	\centering
	\includegraphics[width=3.0in]{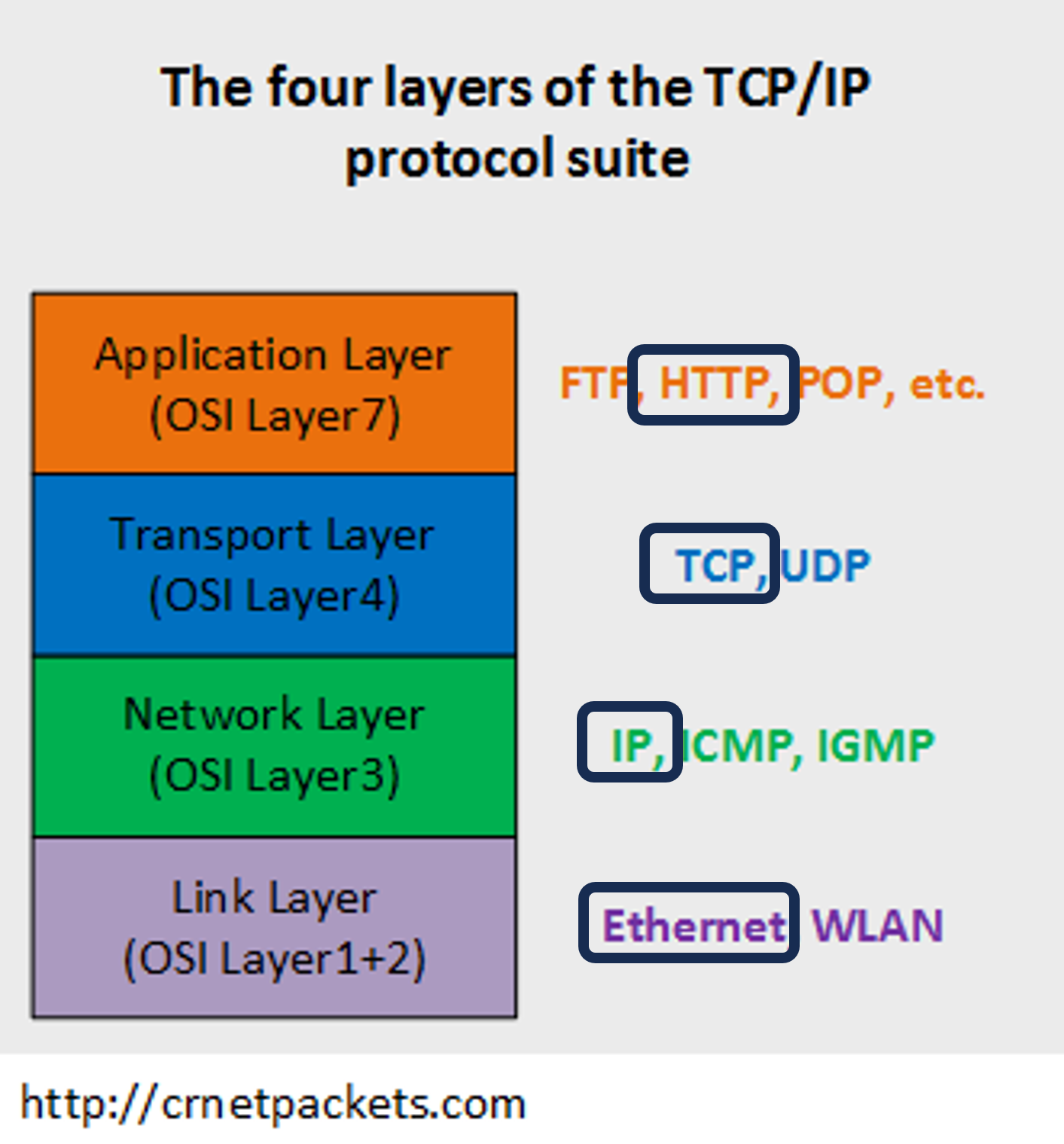}
	\caption{Energy Internet TCP/IP model.}
	\label{fig:TCPIP}
\end{figure}

To achieve free and convenient energy exchange between Energy Internet users without the intervene of the system operator, we in this paper further define Energy Internet protocols, taking Internet protocols as a paradigm. For simplicity, in this paper, we mainly consider a simpler four-layer TCP/IP model as in Fig. \ref{fig:TCPIP}. We believe an extension to the seven-layer OSI model is feasible and is a topic worth future investigation.

\begin{figure}[H]
	\centering
	\includegraphics[width=4.0in]{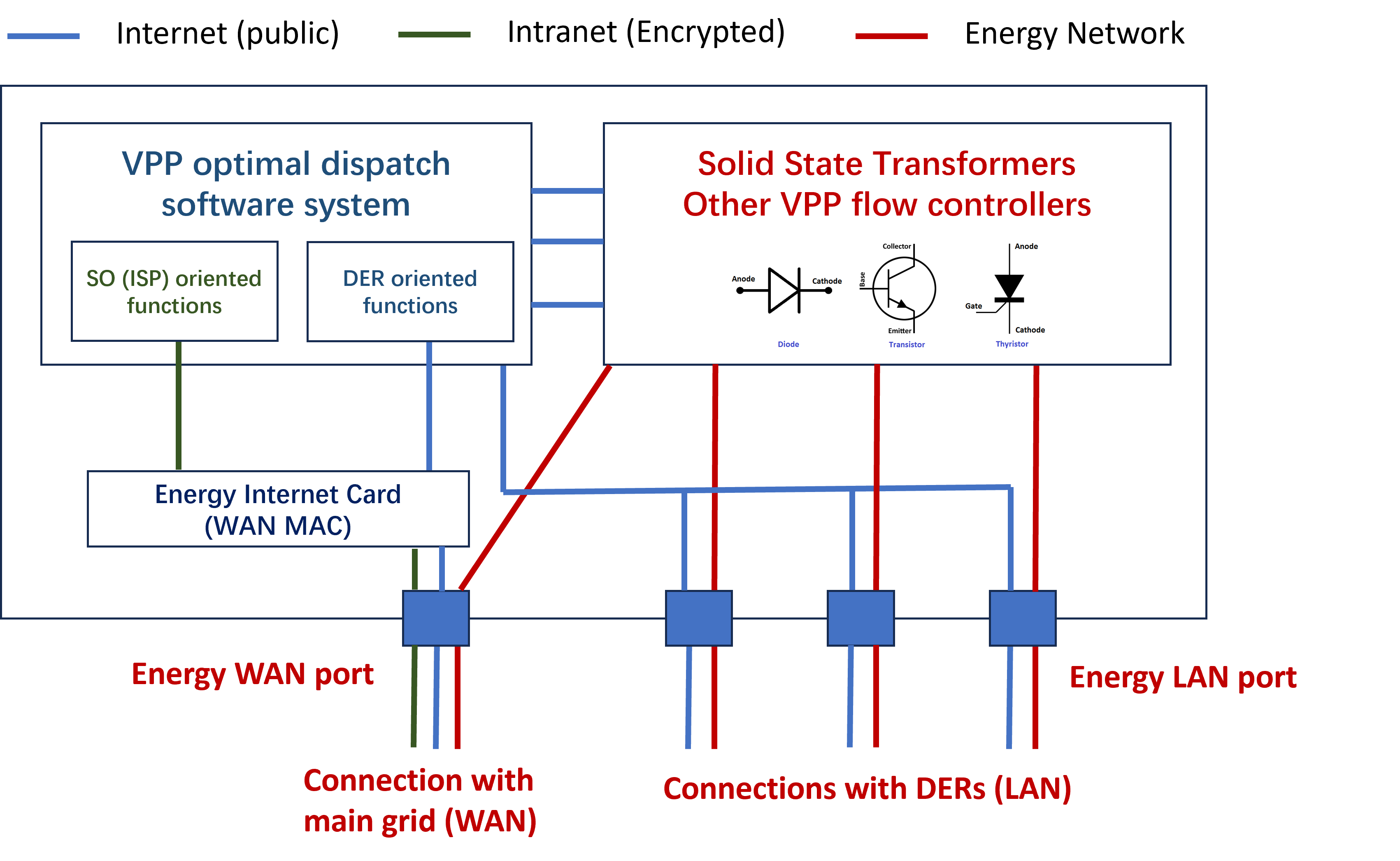}
	\caption{Energy Internet Router.}
	\label{fig:EI}
\end{figure}

\textbf{Link layer}: It establishes physical cable connections in the Internet. For the Energy Internet, connections should be established for both information and energy to enable the transfer of BEE between users. Most of such connections have already been established, while they may not be organized in a standard way. In this paper, we propose that each resource is equipped with an Energy Internet Card, which is also assigned a MAC address as an identification of each resource in the Energy Internet.

\textbf{Network layer}: The most common form is the Internet Protocol (IP). It achieves networking, assigns IP addresses, and realizes basic flow control functions. It will be an essential layer in the Energy Internet as well. Each node/location of the energy network will be assigned with an Energy IP address, which is different than the MAC address for resources. For a mobile energy resource like an electric vehicle or when an energy resource join a new VPP, its Energy IP address changes according but its MAC address will not. User profiles will also go with resources.

The IP header used in the Internet can also be applied to Energy Internet. 

\textbf{Transport layer}: The most common form is the Transmission Control Protocol (TCP). It establishes reliable connections between users in the Internet. In the Energy Internet, most reliability measures can be preserved, and dynamic limits on energy exchange can be implemented to better ensure the security of energy system operations. The TCP header can also be adopted.

\textbf{Application layer}: The dominate form is HTTP. It carries the data communicated in the Internet. In the Energy Internet, similarly, the application layer communicates standard energy between resources. How these energy will be communicated will be decided by Energy Internet Applications, as introduced in the following section.

Note that energy exchange protocols above are not for P2P trading only. They are also implemented for energy exchange between Energy LAN and centralized resources controlled by the Energy ISP.

\section{Energy Internet Applications}

Energy Internet structure and protocols enable reliable and flexible energy exchange between resources without any involvement of a centralized operator. While how energy will be exchanged between resources and how Energy Internet users interact with each other are decided by Energy Internet Applications, similar to the case in the Internet. In this paper, a simple stack-based application is designed. 

We assume there is a pool of waiting BEE sending or asking requests, each with a time frame and price. When a client needs to buy or sell energy, it may choose to browse the list of available BEEs, or it may communicate its exchange request to the platform, who will search the BEE pool for the best match. Remainder quantity will be left in the pool and wait for future BEE. 

We believe better applications will emerge in the future, but their design is out of the scope of this article.

\section{Energy System Operation}

Given all the desirable properties of the Energy Internet, the energy system still needs to operate in a secure and reliable manner. This is the task for the Energy ISP -- current system operators. 

It is shown that if P2P trading will not affect physical nodal power injections -- had they did not trade with each other, they still trade with the grid -- then P2P trading has in fact no impact on the power system operation. This shows the possibility to make the Energy Internet decoupled with the energy system operation.

Namely, the system operator (Energy ISP) will forecast or measure nodal power injections, considering the impact of BEE transfers between clients, and take mainly two types of measures to ensure system security: i) Implement static (in Energy IP) and dynamic (in Energy TCP) quantity limits for energy exchanges, ii) dispatch centralized resources. In particular, most existing functions in system operators are still useful in the second type of measures. When necessary, the Energy ISP may reverse the system operation associated purely with Energy Internet Users' behaviours by artificially adding new exchanges, without being aware of details about BEE transfers between clients. Costs of all measures above, together with losses not captured in BEE transfers, will be reimbursed by Energy Internet Service fees and afforded by all clients in the Energy Internet according to their network usage.

Decoupling between users' behaviours and energy system operations by the Energy ISP also paves the way to address the issue of cyber-security. Energy Internet applications should be deployed in the public domain, where users in the public can freely communicate with each other to decide their BEE transfers. Energy system operation is still executed in a private network, where the Energy ISP dispatches centralized assets and send one-way-out static and dynamic exchange quantity limits through Energy Internet TCP/IP protocols. Users need not and should not know scrutinized information such as power grid topology and parameters, and the Energy ISP need not know details about users' BEE transfer decisions but just need to forecast or measure nodal power injections. These two domains of networks can be physically isolated.

\section{Example}

We compared the proposed Energy Internet framework to traditional centralized dispatch on a four-node toy example. Fig. \ref{fig:Sys} illustrates its configuration:

\begin{figure}[H]
	\centering
	\includegraphics[width=3.0in]{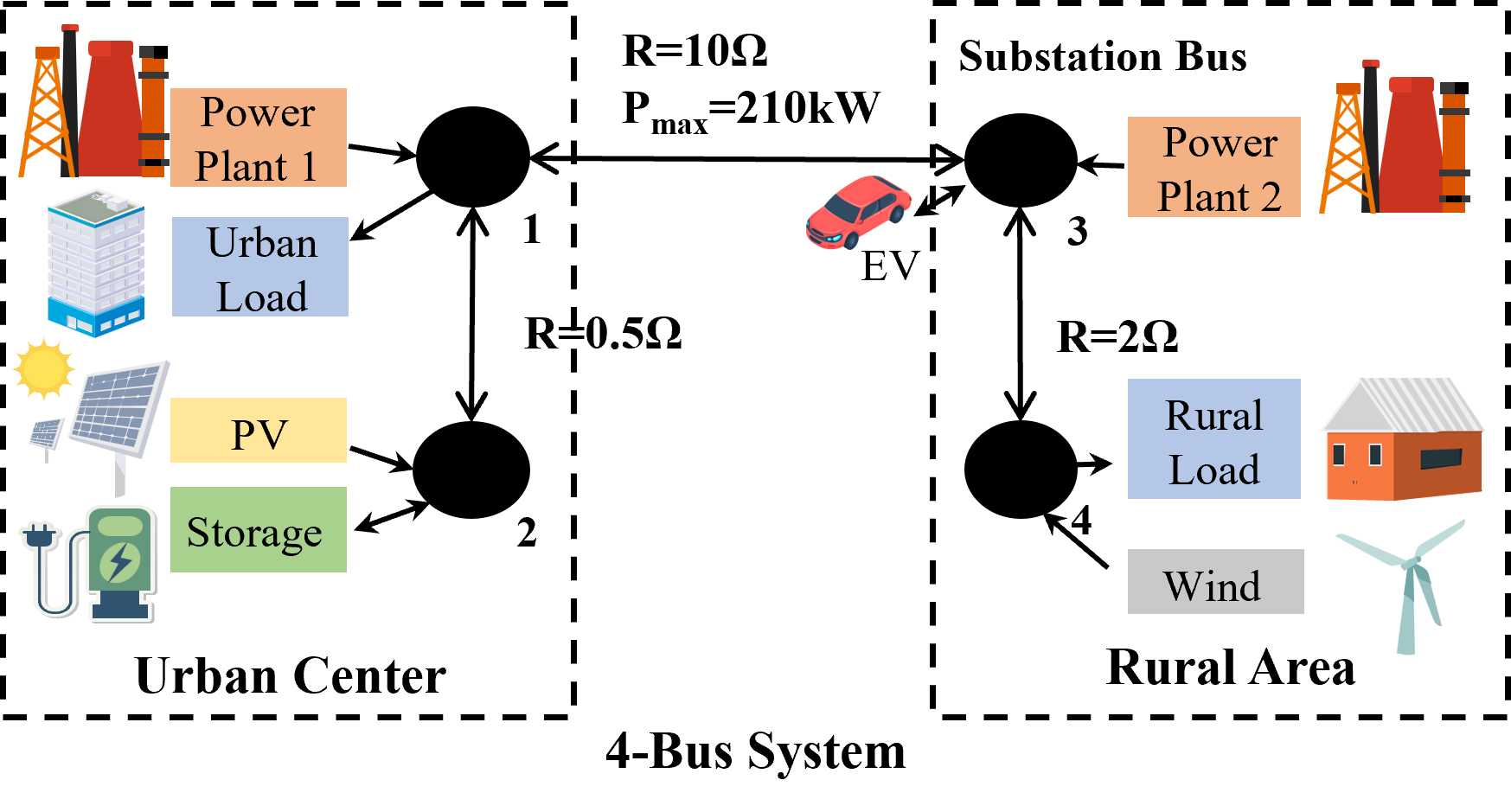}
	\caption{Example System Configuration.}
	\label{fig:Sys}
\end{figure}

\begin{figure}[H]
 \centering
        \vspace{-1cm}
	\includegraphics[width=5.5in]{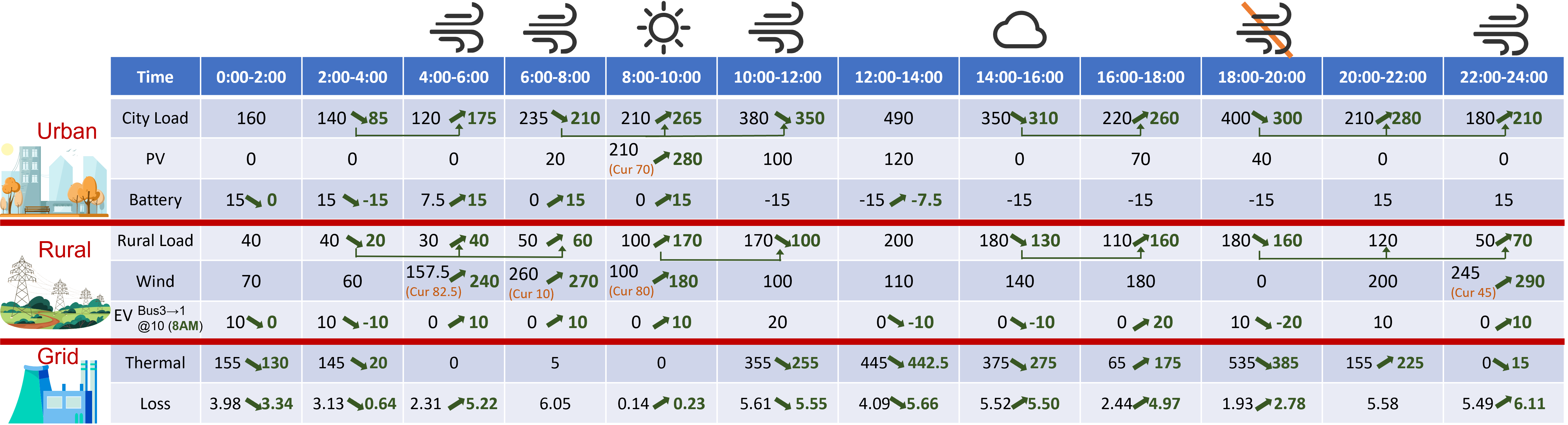}
	\caption{Resources power adjustment.}
	\label{fig:Solution}
\end{figure}

\begin{figure}[H]
	 \centering
        \vspace{-1cm}
        \includegraphics[width=4.5in]{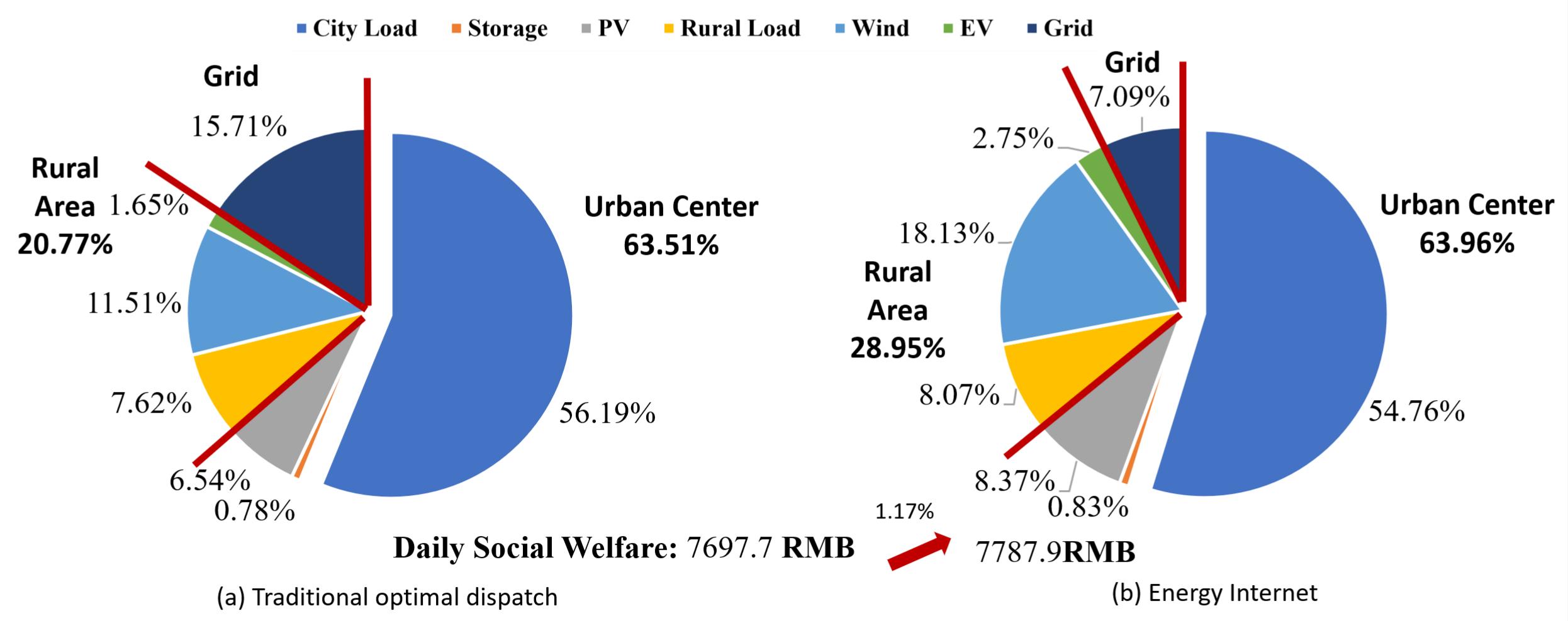}
	\caption{Social welfare and its allocation.}
	\label{fig:Pie}
\end{figure}

\begin{figure}[H]
        \centering
        \vspace{-1cm}
	\includegraphics[width=5in]{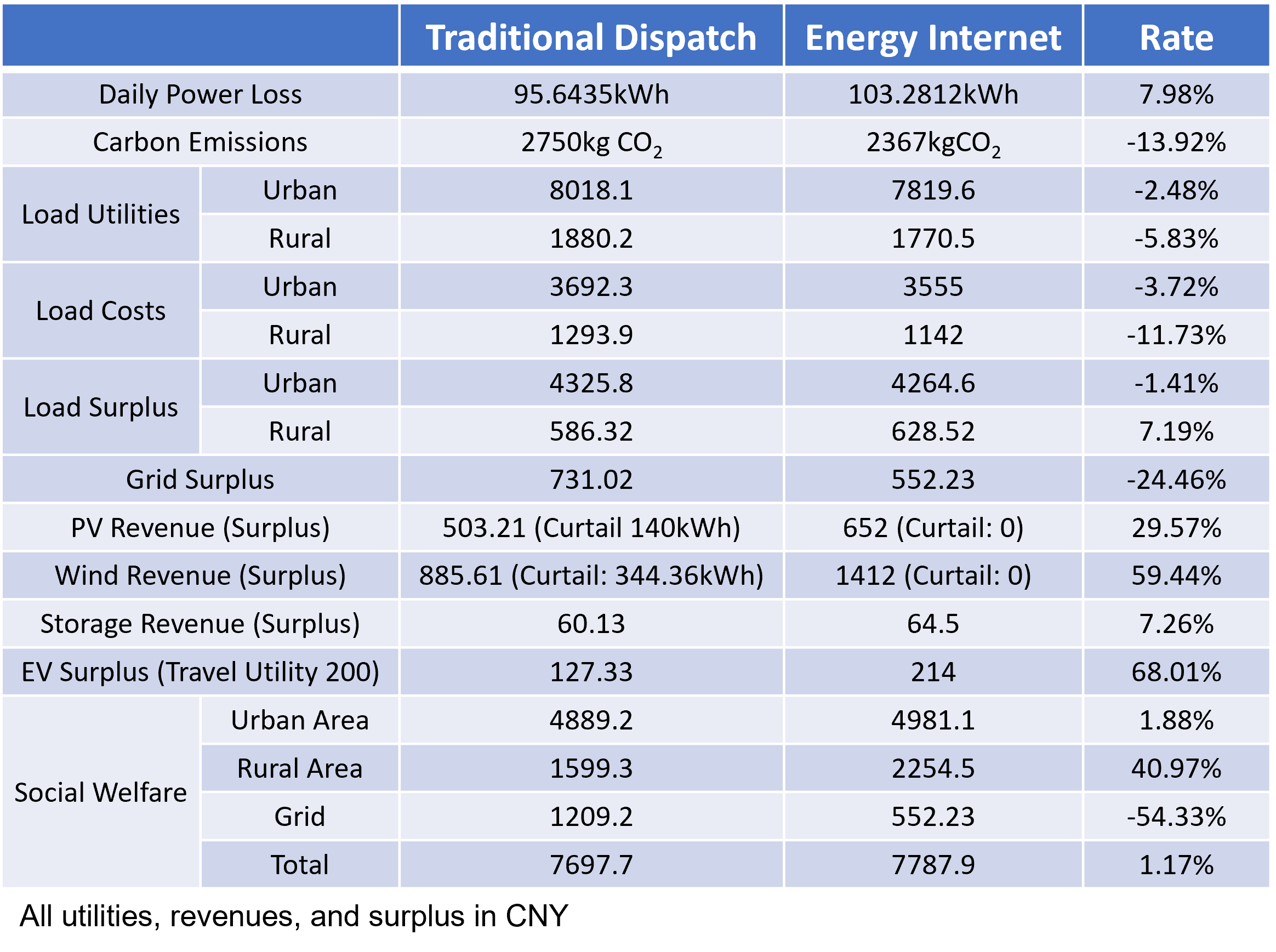}
	\caption{Detailed statistics.}
	\label{fig:Stat}
\end{figure}

The daily operation is split into twelve periods, each representing two hours. Baseline power of these resources and their actions in the Energy Internet are illustrated in Figure \ref{fig:Solution}. The social welfare of the day and its partition are presented in Figure \ref{fig:Pie}, and detailed statistics are given in Figure \ref{fig:Stat}.

\newpage

Compared to traditional operations, the following changes have been observed for Energy Internet in this example:

\begin{itemize}
    \item \textbf{Improved social welfare (1.17\%)}: With resources participating in the optimization process by adjusting their actions. However, the daily power loss slightly increases in this case since resource owners cannot consider system parameters.
    \item \textbf{Significant lower carbon emissions (-13.92\%)}: Energy Internet has eliminated all wind and solar curtailment in this example. Thus, the system is much less reliant on thermal units, and the carbon emission has been effectively reduced.
    \item \textbf{Drastic increase in profits of renewables, battery, and EV}: This can further promote the development of such resources.
    \item \textbf{Surplus shifts from grid to resources}: The grid surplus reduces (-24.46\%) but is still non-negative, consistent with its transition to the Energy ISP. Meanwhile, urban and rural communities have increased surplus ratios. This is indeed the trend we want to see in the process of energy democracy.
    \item \textbf{Better energy justice}: The rural area with a higher load elasticity and excessive wind energy, which is more vulnerable to curtailment, benefits the most. This also demonstrates the Energy Internet's potential to dwindle the energy equity gap. 
\end{itemize}

\section{Conclusions and Discussions}

In summary, this paper develops a tractable design of Energy Internet, emulating the Internet's standardization: with standardized Blocks of Energy Exchange (BEE) as the media, standardized user profiles, Energy Internet Cards that encode and decode BEE transfers to update user profiles, and standardized protocols based on TCP/IP models. Consequently, users of Energy Internet can interact with each other without any interventions from centralized system operators. Operations of the Energy Internet and electric power systems are decoupled. The Energy Internet energy may lead to the systematic reform of current energy ecosystems and the realization of energy democracy, facilitating the goal of carbon neutrality with efforts from not only a few centralized system operators and large power plants but also millions of resources and entities in our society.

It should also be noted that the Internet also has its dark side: Disinformation, insulting content, and even crime platforms. In the proposed design, each resource is assigned a unique MAC address. With the assistance of a credible and attributable distributed ledger system for user profiles, possibly supported by the blockchain, we believe it is possible to hold malicious participants accountable. Nevertheless, this is definitely a complicated issue that is worth further investigation in the future.

\section{Methods}

\subsection{Detailed Settings of the Example}

\textbf{Optimal power flow}: We employ the standard ACOPF model and MATPOWER as the solver to decide the optimal dispatch of thermal power plants from the grid's point of view. However, we do not consider any reactive power-related parameters, e.g., reactive load, branch reactance, or line shunt. The MATPOWER solver will return optimal dispatch of thermal power plants, their total costs, and locational marginal prices $\pi$ for all nodes (differentiated by network loss). 

Note that this step is needed in both the traditional approach and the Energy Internet. For the latter, its ACOPF considers nodal power injections, but not specific P2P transactions, and optimizes the output of power plants considering network loss, which resources without network parameters are unaware of.

\textbf{Load modeling}: We consider elastic load for both urban and rural consumers. When prices increase 10\%, their load power will reduce 5\% and 20\%, respectively. With elasticities above and the quantity-price pair solved from the optimal power flow, we can obtain their demand curve and utility functions. These utility functions are used when calculating consumer surpluses in the traditional approach and Energy Internet.

For the traditional approach, they pay locational marginal prices. In the Energy Internet, consumers may trade with other resources or with the grid. A detailed list of trading patterns will be uploaded in a separate supplemental file. The difference between loads' utilities and payments defines their surpluses.

\textbf{Power Plants}: Two power plants are assumed to have quadratic cost functions $\frac{1}{2}aP^2+bP$, with $a=$0.0008 and 0.0005 and $b=$0.55 CNY/kWh$^2$ and 0.35 CNY/kWh, respectively. Their carbon intensity is set as 0.55tCO$_2/MWh$.


\textbf{Grid Surplus}: In the traditional operation, the grid surplus includes two parts: 1) profits of the two power plants, as the difference between their LMP settled revenue and costs, and 2) merchandise surplus from LMP discrepancies caused by network loss.

In Energy Internet, we propose that as the Energy Internet Service Provider, the grid should at least have a profit rate of 10\%. Namely, its profit, defined by the difference between its revenue from trading with resources and costs solved from OPF, should reach at least 10\% of the latter. Otherwise, resources should pay the deficit proportionally as the Energy Internet Service Fee. In this example, however, the profit rate of the grid is above 10\% without any Energy Internet Service Fee.

\backmatter

\bibliography{refer}


\end{document}